\documentclass[11pt]{article}
\usepackage{moriond}
\usepackage[]{units}
\usepackage[varg]{txfonts}   
\usepackage{multirow}
\usepackage[FIGTOPCAP]{subfigure}
 \usepackage[abs]{overpic}
\usepackage{hyperref}

\usepackage{enumitem}

\def\be{\begin{equation}}
\def\ee{\end{equation}}
\def\bea{\begin{eqnarray}}
\def\eea{\end{eqnarray}}

\begin{document}
\vspace*{4cm}
\title{Measurement of WZ and ZZ production in pp collisions at 8 TeV in final states with b-tagged  jets with the CMS experiment}

\author{ C.~Vernieri on behalf of the CMS collaboration}

\address{Scuola Normale Superiore, Piazza dei Cavalieri~7, Pisa\\
56127, Italy}

\maketitle

\abstract{

In this note we present a measurement of the VZ (V=W,Z) production cross section in proton-proton collisions at $\sqrt{s}=$8 TeV in the VZ$\rightarrow$V$b\bar{b}$ decay mode with V$=$Z$\rightarrow (\nu\bar{\nu}$,$\ell\ell$), V$=$W$\rightarrow \ell{\nu}$, ($\ell=e,\mu$). The results are based on data corresponding to an integrated luminosity of 18.9 fb$^{-1}$ collected with the CMS experiment. The process is observed for the first time in this particular final state with a significance exceeding six standard deviations ($\sigma$). The measured cross sections are consistent with the predictions of NLO calculations. }

\section{Motivations}
The study of VZ (V=W,Z) diboson production in proton-proton ($pp$) collisions provides an important
test of the electroweak sector in the standard model (SM). 
The VZ production in pp collisions at the Large Hadron Collider (LHC) has been measured with CMS and ATLAS
in fully and semi leptonic decay modes ~\cite{CMS-PAS-SMP-12-006,AtlasVV}, while evidence for the VZ $\rightarrow$ V$b\bar{b}$ has been
reported by Tevatron with $4.6$~$\sigma$ significance~\cite{VZTevatron}.
Though the $b\bar{b}$ final state is not sensitive as the leptonic ones, it is interesting being the purest $b\bar{b}$ resonance which allows to test the b-jet identification and reconstruction. 
In addition the VZ final state is the the least reducible background for the VH($b\bar{b}$) search,  its measurement in the relevant phase space for the Higgs boson search represents a strong validation of the VH($b\bar{b}$) analysis strategy~\cite{VHbb}. 

\section{From VH to VZ}
At LHC the most sensitive search for the Higgs boson decaying into a b quark pair is performed requiring the presence of an associated vector boson V decaying leptonically, to suppress the multi-jet QCD contribution and to provide an efficient trigger path (charged leptons, E$^{\mathrm{miss}}_T$). According to the associated V decay mode, the corresponding analysis channels are the 0-lepton, 1-lepton and 2-leptons, where electron and muon channels are analyzed separately.\\
Given the p$_{T}$(V)-dependency of the cross section, the VH($b\bar{b}$) search requires a large boost ($\sim$~100 GeV) for both the $b\bar{b}$ pair and the V boson to reduce the Z/W+jets and $t\bar{t}$ backgrounds. For each channel, several p$_T$(V) bins with different signal and background contributions are analyzed separately in order to optimize the significance, see Table~\ref{tab:boost}.  
The VZ measurement is then performed in the phase space defined by p$_{T}$(V)$>$100 GeV, where approximately 15\% of the WZ and 14\% of the ZZ total inclusive cross sections are contained.

\begin{table}[h!]    
\centering    
%
\begin{tabular}{c | c c c }

\hline\hline
\footnotesize{{p$_T$(V) bin }} &\footnotesize{ low}& \footnotesize{intermediate} & \footnotesize{high}   \\
\hline               
\footnotesize{0-lep}   &   \footnotesize{[100,130]} &\footnotesize{[130,170]} & \footnotesize{$>170$}\\
\footnotesize{1-lep} &   \footnotesize{[100,130]} & \footnotesize{[130,180]} & \footnotesize{$>180$}\\
\footnotesize{2-lep}  &  \footnotesize{}&&\footnotesize{$>100$}\\
\hline 
\hline   

\end{tabular}

\caption{p$_T$(V) category definition for each decay channel. Values are in [GeV].}
\label{tab:boost}
%
%
%
\end{table}

\section{Analysis Strategy}
\label{sec-1}
The event selection is based on the reconstruction of the leptonic decay of the V and of the Z boson decay into two b-tagged jets. 
Dominant backgrounds to VZ production originate from V+ jets,
$t\bar{t}$, single top, multi-jet (QCD) and VH production.
Besides requiring requiring a large boost of the vector boson and the $b\bar{b}$ pair~\cite{sub}, also the back-to-back topology and minimal additional jet activity requirements are used to reduce $t\bar{t}$ and single top contributions, while M($b\bar{b}$) helps to separate VZ from V+jets and VH. \\
The reconstruction of the Z $\rightarrow b \bar{b}$ decay is made by selecting the pair of jets in the event with the highest p$_T$(jj). The two jets are tagged as b-jets using the Combined Secondary Vertex  (CSV) algorithm~\cite{CSV}. 
W$\rightarrow\ell\nu$ decays are identified by requiring a single isolated lepton and additional  E$^{\mathrm{miss}}_T$.  The identification of Z$\rightarrow \nu\nu$ decays requires the E$^{\mathrm{miss}}_T$ and no isolated leptons in the event. Z$\rightarrow\ell\ell$ candidates are reconstructed by combining isolated, oppositely charged pairs of electrons or muons and requiring the dilepton invariant mass to be within a Z mass window of 30 GeV width.  Detailed event selection is reported in~\cite{VZbbCMS}.\\
The analysis strategy main features are: 
\begin{enumerate}[topsep=0pt,itemsep=-1ex,partopsep=1ex,parsep=1ex]
\item Extraction of the normalization of the dominant backgrounds from the data 
\item Improvement of the M($b\bar{b}$) resolution with b-jet energy specific corrections (regression)
\item More efficient discrimination of the signal through the use of a multivariate analysis.
\end{enumerate}

\subsection{Background Estimate}
Control regions kinematically close to the signal region are identified in data and used to correct the Monte Carlo yields estimated for the main background processes:  V+jets (split into 0/1/2b content at generator level) and $t\bar{t}$ production. 
A set of simultaneous fits is performed to several distributions of discriminating variables in the control regions, separately in each channel, to obtain consistent scale factors (SF) by which the Monte
Carlo yields are adjusted.  These scale factors account not only for cross section discrepancies in the theoretical predictions, but also for potential residual differences in physics object selection in the boosted phase space.
Good agreement between data and simulation is found after applying the fitted SF in several control regions for all modes as shown by the dijet invariant mass distribution for the combination of all five channels, in all p$_T$(V) bins reported in Fig.~\ref{fig:MJJ-combined}, (i).

\subsection{Specific corrections for b-jet energy}
An optimal $b\bar{b}$ invariant mass resolution is important to separate VZ and VH processes. 
The M($b\bar{b}$) resolution is already improved by selecting the boosted phase space. 
A b-jet specific correction is derived in this analysis in an attempt to recalibrate to the true b-quark energy, by applying multivariate regression techniques similar to those used by the CDF experiment~\cite{CDF}. 
The regression is essentially a multi-dimensional calibration to the particle level - including neutrinos - which exploits all the relevant properties of a b-quark jet.
This procedure also addresses naturally the problem of semi-leptonic b decays (35\%), which results in lower response with respect to the light quark/gluon induced jets. 
A specialized BDT \cite{tmva} is trained on simulated signal events and provides a correction factor that improves both the b-jet energy measurement and its resolution. \\
Inputs are chosen among variables that are correlated with the b-quark energy and well measured. They include detailed jet structure information about tracks and jet constituents which differs from light flavor quarks/gluons jets. 
Information from B-hadron decays on the reconstructed secondary vertices are used as well as soft lepton from semileptonic decay when available, providing an independent estimate of the b quark p$_T$. 
For the Z($\ell\ell$) channel is exploited also the information carried by the variables related to the E$^{\mathrm{miss}}_T$ vector. Since there is no real E$^{\mathrm{miss}}_T$ in the event it works as a kinematic constraints for the momentum equilibrium in the transverse plane.

\begin{figure}[ht]
\centering

	\begin{overpic}[width=0.33\textwidth]{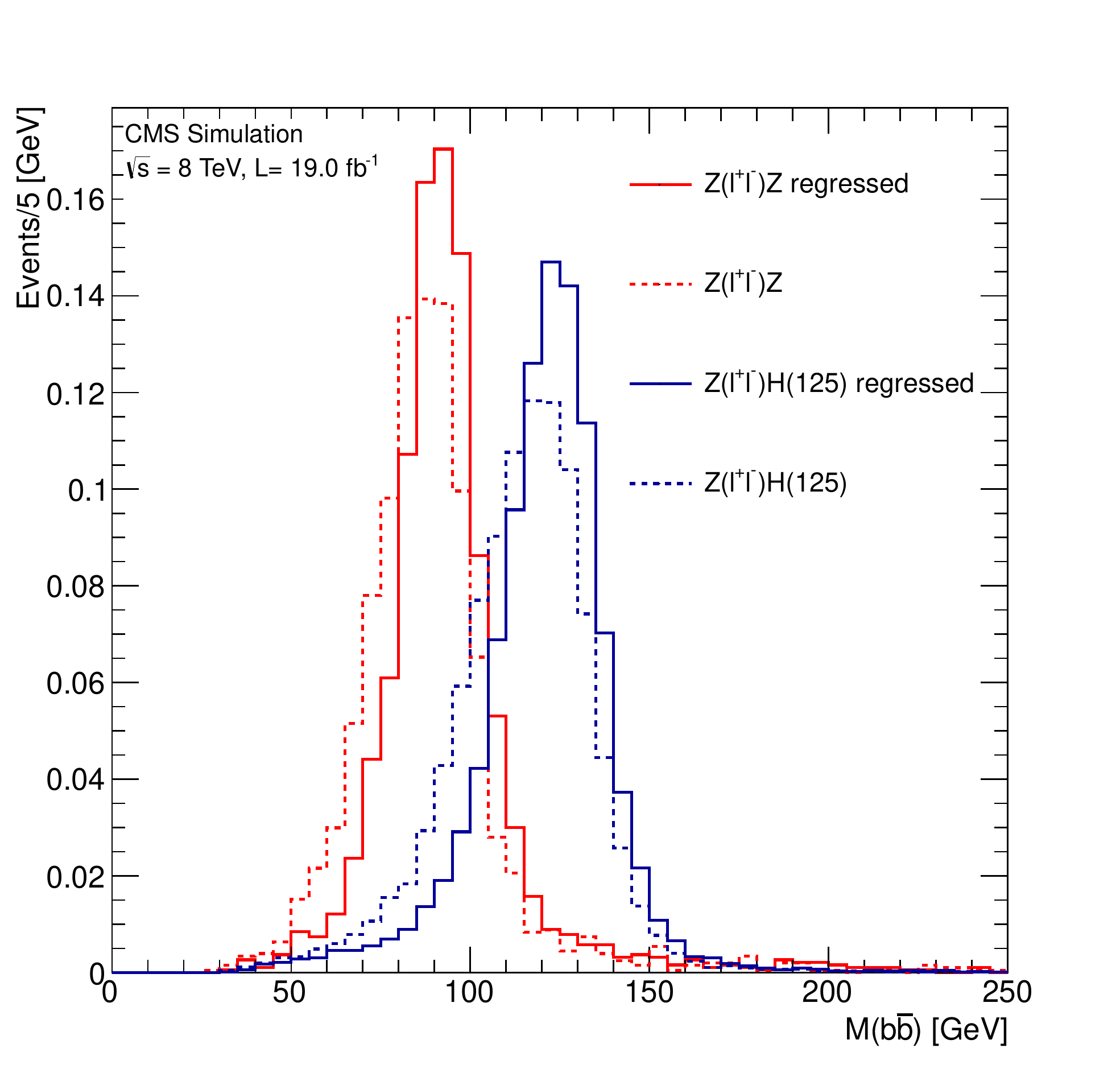}
	\put(22,100){(i)}
\end{overpic}
\begin{overpic}[width=0.31\textwidth]{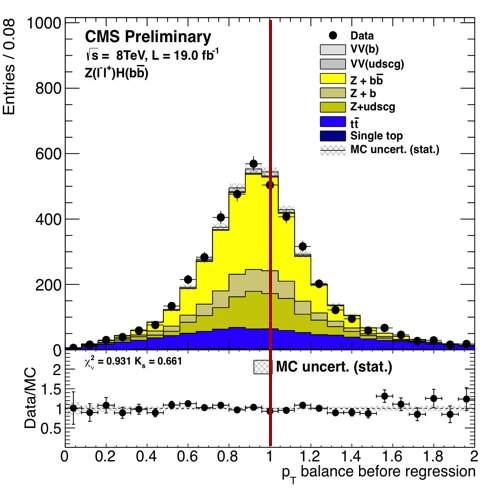}
	\put(30,100){(ii)}
\end{overpic}
\begin{overpic}[width=0.31\textwidth]{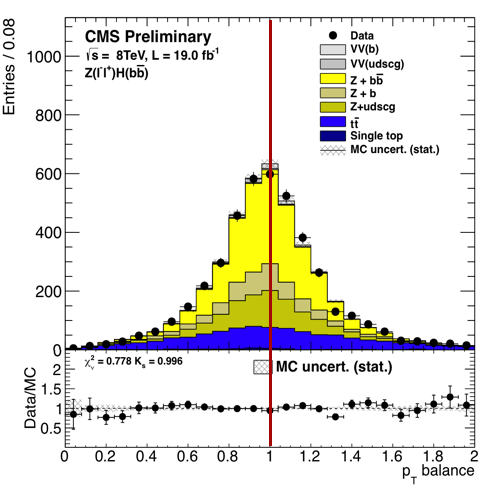}
	\put(30,100){(iii)}
\end{overpic}

\caption{(i) Mass difference between ZH and ZZ simulated processed before and after the regression is applied. Distribution of the p$_T$ balance on data versus MC before (ii) and after (iii) the regression.}
\label{fig:Validation}
\end{figure}

The improvement on the M($b\bar{b}$) resolution when the corrected jet energies are used is approximately 15\%, impacting the analysis sensitivity by
10--20\%, depending on the specific channel. After correction the resolution of the Z$\rightarrow b\bar{b}$ di-jet invariant mass is about 10\% and it results in a better separation of the VZ/VH processes as reported in Fig.~\ref{fig:Validation} (i).

\begin{figure}[htbp]
 \begin{center}
	

    \begin{overpic}[width=0.32\textwidth]{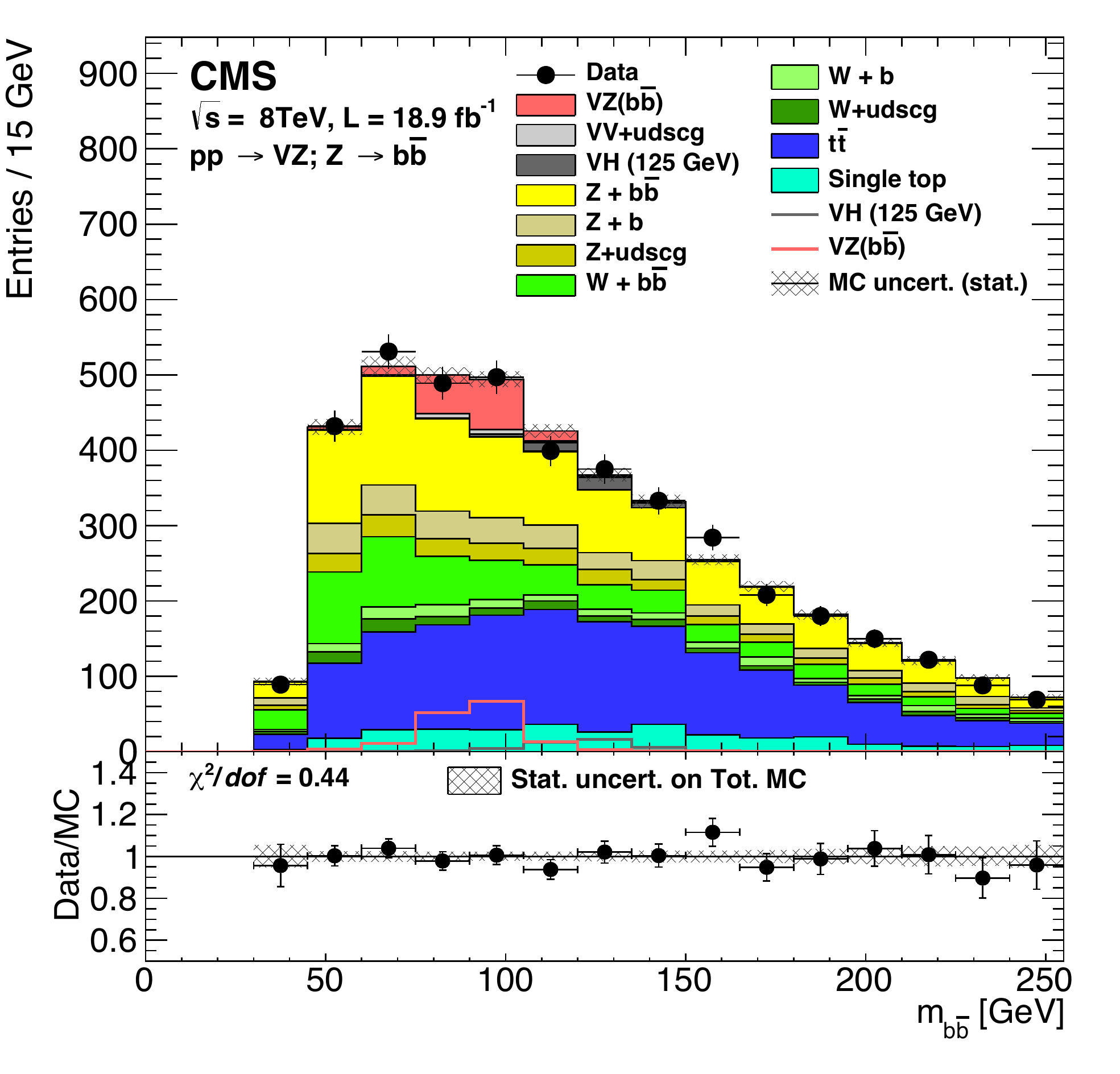}
	\put(25,100){(i)}
\end{overpic}
    \begin{overpic}[width=0.32\textwidth]{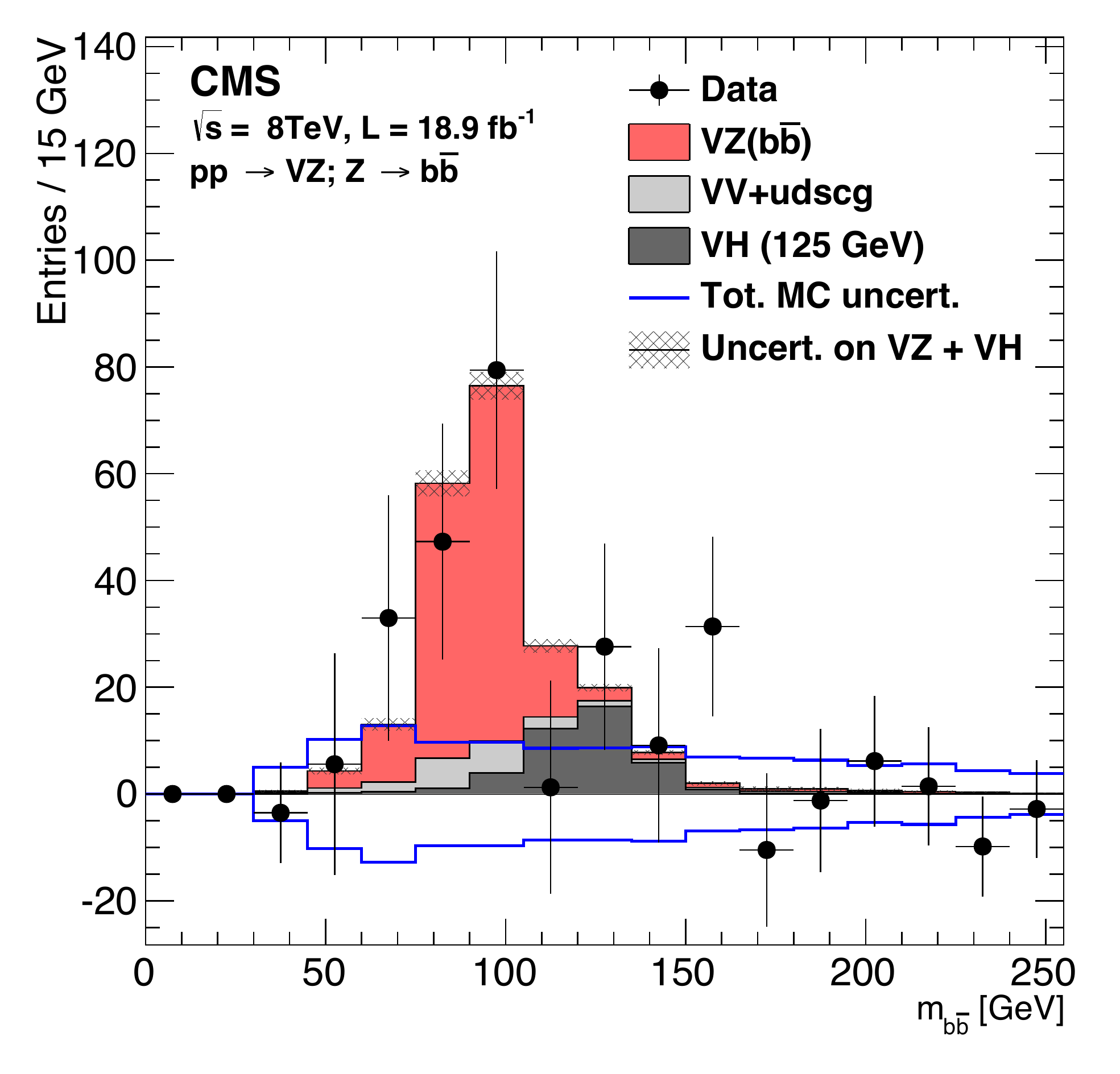}
	\put(25,100){(ii)}
\end{overpic}
     \begin{overpic}[width=0.32\textwidth]{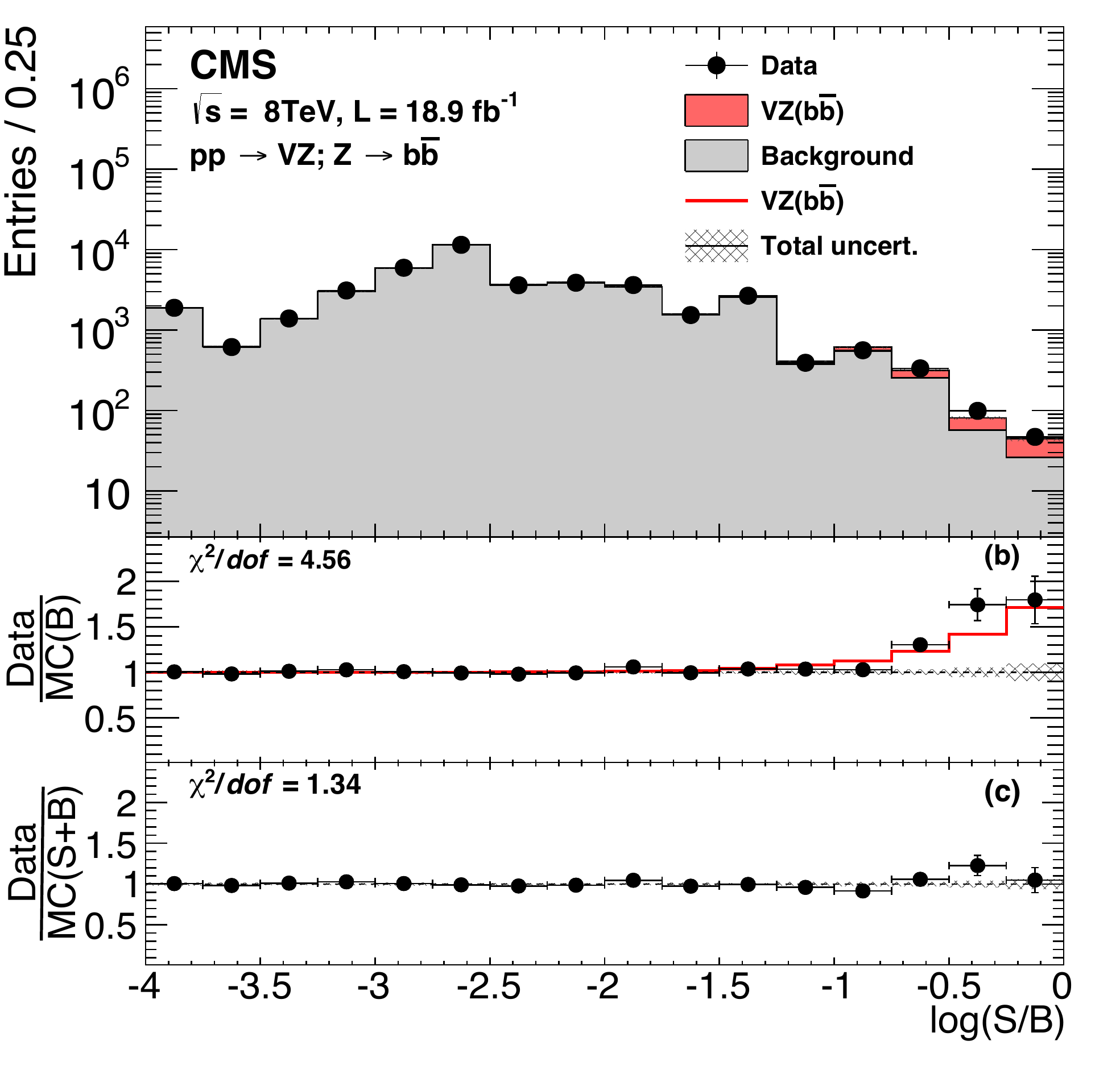}
	\put(120,100){(iii)}
\end{overpic}

 \end{center}
 \caption{(i) Dijet invariant mass distribution. (ii) Same distribution after the non resonant subtraction.  (iii) Events are sorted in bins of similar expected S/B as given by the BDT discriminant output value.}
   \label{fig:MJJ-combined}

\end{figure}

The regression technique has been validated also looking at the p$_T$ balance between the di-lepton and the di-jet systems

distribution in a Z($\ell\ell$)$+b\bar{b}$ enriched data sample ($\ell=e, \mu$). Exactly two central jets are required and then we use CSV to select $b\bar{b}$ candidates. Candidate Z($\ell\ell$) decays are reconstructed requiring $75<M(\ell\ell)<105$ GeV.
A data to simulation comparison is shown in Fig.~\ref{fig:Validation} before regression (ii) and after regression (iii), reporting an improved resolution and scale. The mean of the distribution is centered at unity after
the regression and the resolution, defined as the sigma divided by the mean, (sigma and mean are evaluated through a fit with a Gaussian function) is improved by $20\%$. Fig.~\ref{fig:MJJ-combined} (ii) shows the M($b\bar{b}$) distribution after the non resonant backgrounds subtraction. The VZ signal is clearly
visible with a yield compatible with the SM expectation and a significance of 4.1$\sigma$.


\subsection{Multivariate Analysis}
A more efficient separation of the signal from the background is achieved by exploiting together with the improved M($b\bar{b}$) other properties of the signal signature through a multivariate analysis.
A BDT discriminant is trained using simulated samples for signal and all background processes. Among the most discriminating variables for all channels are the dijet invariant mass distribution, M($b\bar{b}$), the number of additional jets, $N_{\mathrm{aj}}$, the value of CSV for the jets and the distance in $\eta$--$\phi$ between the di-jet system.\\
The signal and background yields are then estimated in a fit region defined in the continuous output of the BDT, and a shape analysis is performed on that output.\\
Fig.~\ref{fig:MJJ-combined} (c)
combines all discriminants into a single distribution where all
events, for all channels, are sorted in bins of similar expected signal-to-background
ratio, as given by the value of the output of their corresponding BDT
discriminants. The
observed excess of events in the bins with the largest
signal-to-background ratio is consistent with what is expected from
VZ production. The VZ process is observed with a statistical significance of
6.3$\sigma$ (5.9~$\sigma$ expected). This corresponds
to a signal strength relative to the SM of $\mu=1.09^{+0.24}_{-0.21}$.

\section{Results}
\begin{figure}
\centering
\begin{minipage}{0.4\linewidth}

\centerline{\includegraphics[angle=0,width=0.9\linewidth]{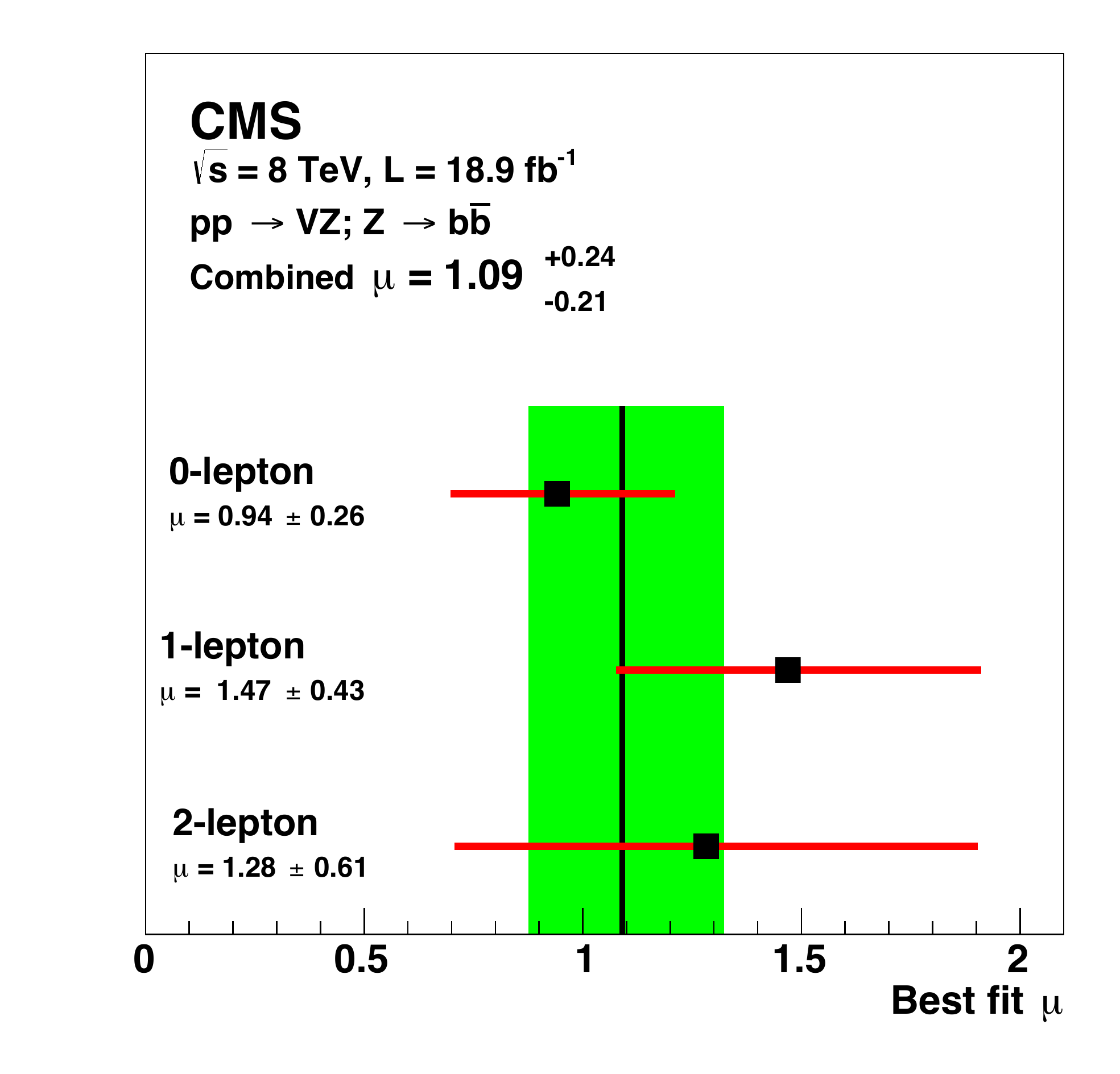}}
\end{minipage}
\begin{minipage}{0.4\linewidth}
\centerline{\includegraphics[angle=0,width=0.9\linewidth]{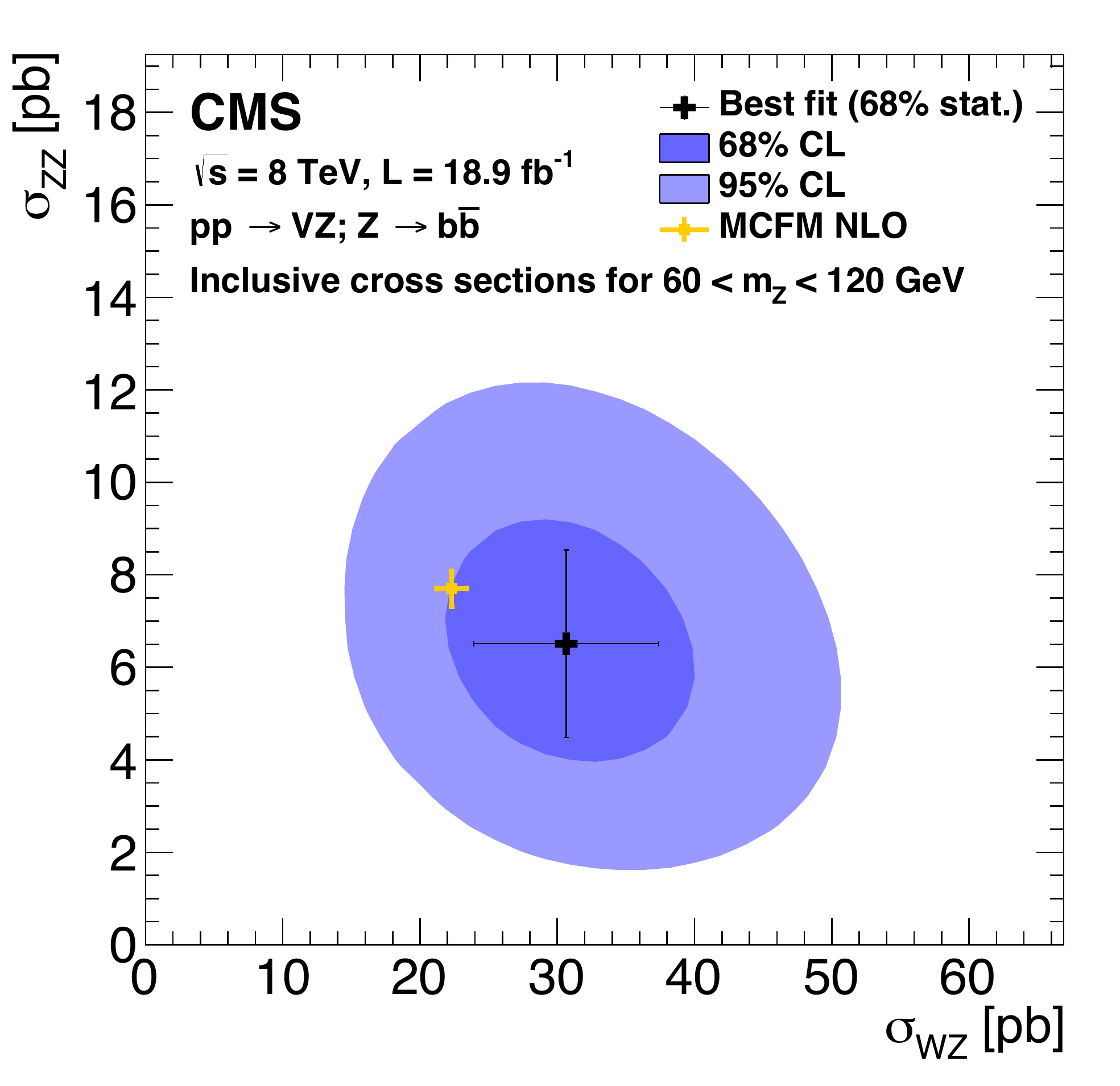}}
\end{minipage}
\caption{(a) The best-fit value of the VZ production cross section,
relative to the SM cross section, for partial
combination of channels and for all channels combined (band). (b) The 68\%/95\% CL contour regions for the WZ-ZZ production cross-sections.}
   \label{fig:SignalStrength}
\end{figure}

The total cross sections are determined by including all final states
in a simultaneous constrained fit on the number of observed events in all categories.
We extract the best fit value relative to the SM expectation at NLO.
All cross section (relative to the SM value) extracted for individual channels
provide compatible values with each other and the SM expectation
Fig.~\ref{fig:SignalStrength} (a). To extract the WZ and ZZ
cross-section a simultaneous fit floating both contributions
independently is performed (Fig.~\ref{fig:SignalStrength}-b).
The best-fit is found to be at ${\mu_{\mathrm{WZ}} = 1.37 {}_{-0.37}^{+0.42}}$
and ${\mu_{\mathrm{ZZ}} = 0.85 {}_{-0.31}^{+0.34}}$.\\
The resulting cross sections are measured to be:\\\vspace{-3mm}
$${\sigma (pp \to \mathrm{WZ}) = 30.7 \pm 9.3(\mathrm{stat.}) \pm 7.1 (\mathrm{syst.}) \pm 4.1 (\mathrm{theo.}) \pm 1.0 (\mathrm{lumi.})\, \rm{pb}}$$\vspace{-5mm}
$${\sigma (pp \to \mathrm{ZZ})  = 6.5 \pm 1.7(\mathrm{stat.}) \pm 1.0 (\mathrm{syst.}) \pm 0.9 (\mathrm{theo.}) \pm 0.2 (\mathrm{lumi.})\, \rm{pb}}$$
for the Z boson produced in the mass region $60<M_Z<120$~GeV, the measurements are found in agreement with the NLO MCFM predictions.

\end{document}